\documentclass[a4paper]{article}

\usepackage{times}          
\usepackage[utf8]{inputenc} 
\usepackage[T1]{fontenc}    
\usepackage[english]{babel}
\usepackage{amssymb, amsmath}
\usepackage{url}            
\usepackage{relsize}        
\usepackage{booktabs,arydshln}       
\usepackage{amsfonts}       
\usepackage{nicefrac}       
\usepackage{microtype}      
\usepackage{pifont}         
\usepackage[pdftex]{graphicx}
\usepackage[numbers,sort,compress]{natbib}
\usepackage{hyperref}       
\usepackage[accepted]{icml2017}

\begin{document}

\twocolumn[
\icmltitle{Interactive Exploration and Discovery of Scientific Publications with PubVis}

\begin{icmlauthorlist}
\icmlauthor{Franziska Horn}{}\\
Machine Learning Group, Technische Universität Berlin, Germany\\
  \texttt{cod3licious@gmail.com}
\end{icmlauthorlist}

\icmlkeywords{pubvis, webapp, interactive, data visualization, information extraction}

\vskip 0.3in
]

\begin{abstract}
With an exponentially growing number of scientific papers published each year, advanced tools for exploring and discovering publications of interest are becoming indispensable. To empower users beyond a simple keyword search provided e.g.~by Google Scholar, we present the novel web application \emph{PubVis}. Powered by a variety of machine learning techniques, it combines essential features to help researchers find the content most relevant to them. An interactive visualization of a large collection of scientific publications provides an overview of the field and encourages the user to explore articles beyond a narrow research focus. This is augmented by personalized content based article recommendations as well as an advanced full text search to discover relevant references. The open sourced implementation of the app can be easily set up and run locally on a desktop computer to provide access to content tailored to the specific needs of individual users. Additionally, a PubVis demo with access to a collection of $10,000$ papers can be tested online.
\end{abstract}

\section{Introduction}
The Web of Science\footnote{by Thomson Reuters (\url{http://wokinfo.com/})} has a record of almost 60 million published scientific papers from all areas and, concerned with biomedical literature specifically, the PubMed database comprises more than 26 million citations from MEDLINE, life science journals, and online books.\footnote{\url{https://www.ncbi.nlm.nih.gov/pubmed/}} This is already an incomprehensibly large amount of published literature and there is an exponential increase in the number of articles published per year~\cite{bjork2008global}.

With many subfields and fragmented communities, getting an overview of current and accumulated research can be quite challenging and time consuming, especially if you do not know where to start. While review papers and conferences aim to get researchers up to speed on the developments in the field, they can often only provide a biased view on the subject matter, tainted by current trends and personal preferences of the authors and reviewers. Generally, researchers have to rely on simple keyword searches, e.g.~as implemented by Google Scholar, to obtain a specific piece of information. While being easy to use and widely applicable, keyword searches can only comb through the texts to match a query, thereby only scratching the surface of the semantic relations encoded in the unstructured data. While Google does a very good job of delivering the papers you searched for, what if you do not yet know what you are looking for? What if you are new to the field or just want to explore the research beyond your field of expertise? It is hard to search for keywords you are not even aware exist. 

While content discovery and recommendation systems are widely-used in commercial settings for movies \cite{bell2007lessons} or news \cite{li2010contextual}, for example, comparatively few solutions are available for scientific publications. Besides the fact that many of the existing services seem to be poorly maintained after their initial publication, they often rely on user ratings \cite{yoneya2007pure,wang2011collaborative} or citations \cite{gipp2009scienstein}, which can again lead to biased recommendations. The Science Concierge \cite{achakulvisut2016science} computes the similarities between papers' abstracts to provide content based recommendations, however it still requires the user to initially search for articles of interest. As a promising and well maintained project, the Arxiv Sanity Preserver by Andrej Karpathy\footnote{\url{http://www.arxiv-sanity.com/}} lists recent arXiv preprints and enables the user to explore similar papers as well as get personalized recommendations generated based on papers of interest. However, none of these existing solutions provides the user with a global overview of a field to put relevant papers into context.

We present the novel web application \emph{PubVis}\footnote{\url{http://pubvis.herokuapp.com/}\\and \url{http://arxvis.herokuapp.com/}}, which combines a set of features to help researchers discover the content most relevant to them. An interactive visualization of a large collection of scientific publications provides the initial overview of a field, including clusters of subtopics, and invites users to also explore articles from outside their research focus. Besides a simple keyword search to find articles of interest, it is possible to upload the entire abstract of a current paper draft to search for related articles, e.g.~to ensure you are not missing recently published key references. Additionally, personalized article recommendations are available based on the content of other papers the user has marked as relevant.

In the following sections, the individual features of the web application are described in more detail with a special focus on the machine learning techniques behind them. This includes obtaining an exemplary dataset of PubMed abstracts about various cancer types (Section~\ref{sec:dataset}), computing similarities between articles (Section~\ref{sec:similar}), creating an interactive visualization of the dataset (Section~\ref{sec:viz}), searching for related articles based on content similarity (Section~\ref{sec:search}), and generating personalized article recommendations (Section~\ref{sec:rec}). The paper concludes with a discussion and outlook.

\section{Overview of the app}
The PubVis web application is implemented in Python using the Flask microframework to build a REST API back-end delivering all the content in JSON format. On top, the front-end displays the retrieved data, relying heavily on the d3.js library to generate the interactive visualization. The app can be run locally on a desktop computer (with a reasonable amount of RAM) and includes all the code necessary to download paper abstracts from the web and prepare this data for exploration and discovery.\footnote{\url{https://github.com/cod3licious/pubvis}} 

For the initial setup of the app as well as for updating it with new content, a series of actions has to be performed, detailed in the next sections, to scrape new content, update the article similarities and search index, and to compute the embedding coordinates used for the visualization. To save expensive resources when deploying the app to a cloud based service such as Heroku, all updates of the data can still be executed from a local machine by simply connecting to the database running in the cloud. 

Two demos of the app can be tested online, one, discussed in more detail below, includes around $10,000$ PubMed abstracts about different cancer types,\footnote{\url{http://pubvis.herokuapp.com/}} while the other provides access to $10,000$ recent arXiv preprints from the area of machine learning.\footnote{\url{http://arxvis.herokuapp.com/}}

\section{Obtaining data}\label{sec:dataset}
\begin{figure}[ht!]
  \centering
      \includegraphics[width=0.45\textwidth]{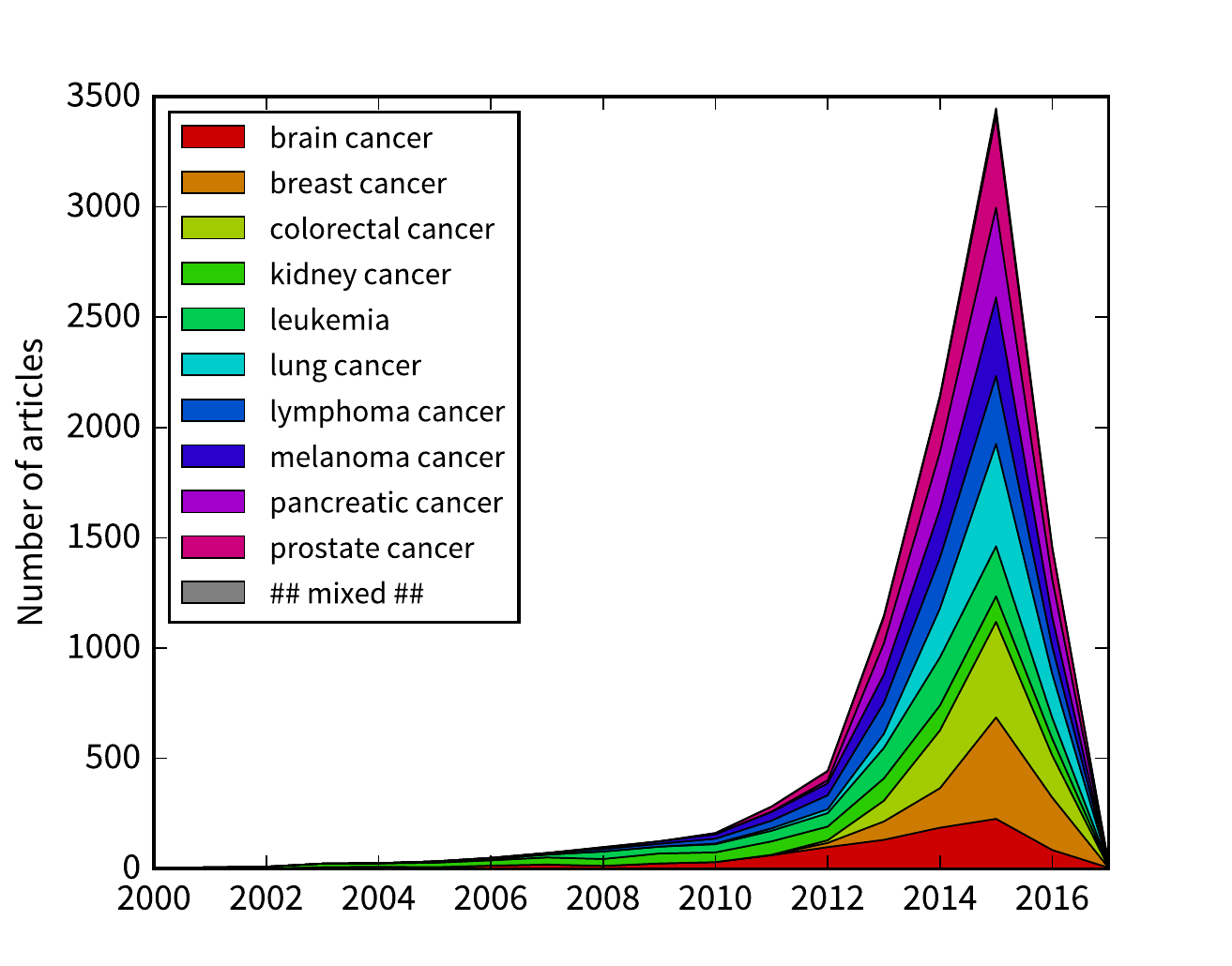}
  \caption{Number of paper abstracts obtained in February 2017 for every cancer type by year.}
  \label{fig:cancerdataset}
\end{figure}

As approximately 40\% of all men and women are diagnosed with a form of cancer during their lifetime~\cite{howlander2013seer}, it is no surprise that cancer is one of the most researched biological topics with many research papers publicly available on PubMed. Using the PubMed API\footnote{\url{http://www.ncbi.nlm.nih.gov/books/NBK25500/}} we create an exemplary scientific literature dataset containing almost $10,000$ abstracts associated with ten different cancer types (Figure~\ref{fig:cancerdataset}). 
While all following results are based on this dataset, it is also easy to populate PubVis with different papers. By default, the app includes functions to obtain abstracts from both the PubMed and arXiv API, and it is possible to add content from other sources to PubVis by writing a custom scraper.

\section{Similar publications}\label{sec:similar}
\begin{figure*}
  \centering
      \includegraphics[width=0.95\textwidth]{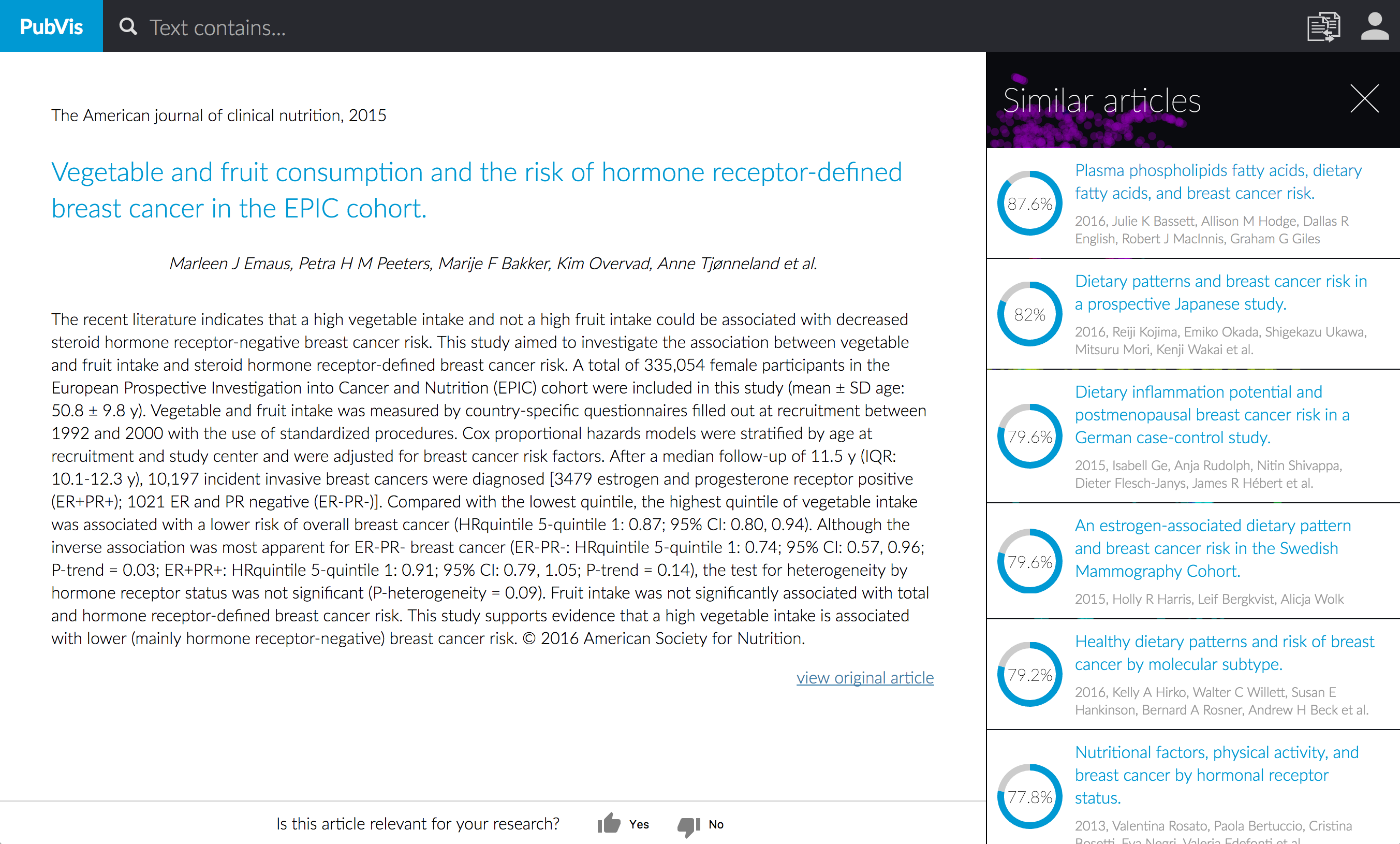}
  \caption{Article view with similar articles.}
  \label{fig:similar}
\end{figure*}

For every article in PubVis, related papers are suggested based on the similarities of the articles' content (Figure~\ref{fig:similar}). 
To compute these similarities, the papers first have to be encoded as numerical feature vectors. By computing weighted counts of all words in a paper's text, a \emph{tf-idf} bag-of-words (BOW) representation is generated for each article. \emph{Tf} stands for `term frequency', which is computed for every document and term by counting how often a word occurs in the text. Some words, such as `the' or `and', occur frequently in almost all documents, but are not very descriptive. Their influence can be reduced by weighting the term frequencies with their inverse document frequency (\emph{idf}). The \emph{idf} of a term $t$ is calculated as the logarithm of the total number of documents in the corpus, $|D|$, divided by the number of documents which contain term $t$, i.e.
\begin{align*}
 \text{idf}\,(t) &= \log {|D|\over |\{d \in D\text{ : }t \in d\}|}.
\end{align*}
The resulting sparse, high dimensional BOW representations of the documents are then transformed using latent semantic analysis (LSA). LSA is a simple topic modeling technique used to reduce noise and create more overlap between document vectors from similar topics. To this end, the dimensionality of the BOW vectors is reduced to a fixed number of components using singular value decomposition.
Finally, the similarity between two documents is computed as the cosine similarity of their LSA vectors.

\section{Interactive visualization}\label{sec:viz}
To create the interactive visualization (Figure~\ref{fig:viz} in the Appendix), the heart of PubVis, the papers' LSA vectors are embedded in two dimensions using t-SNE \cite{van2008visualizing}. The algorithm's ability to preserve local neighborhoods in the embedding makes it an excellent choice for creating a visualization that can be explored to discover related articles. 

To obtain the low dimensional coordinates for a set of data points, t-SNE first constructs a probability distribution over pairs of high-dimensional input data points based on their euclidean distance in the original space. A similar probability distribution is defined over the pairwise distances in the low dimensional embedding space and then the optimal solution is obtained by minimizing the KL-divergence between both distributions iteratively using gradient descent until a local minimum is reached.
\begin{figure*}
  \centering
      \includegraphics[width=\textwidth]{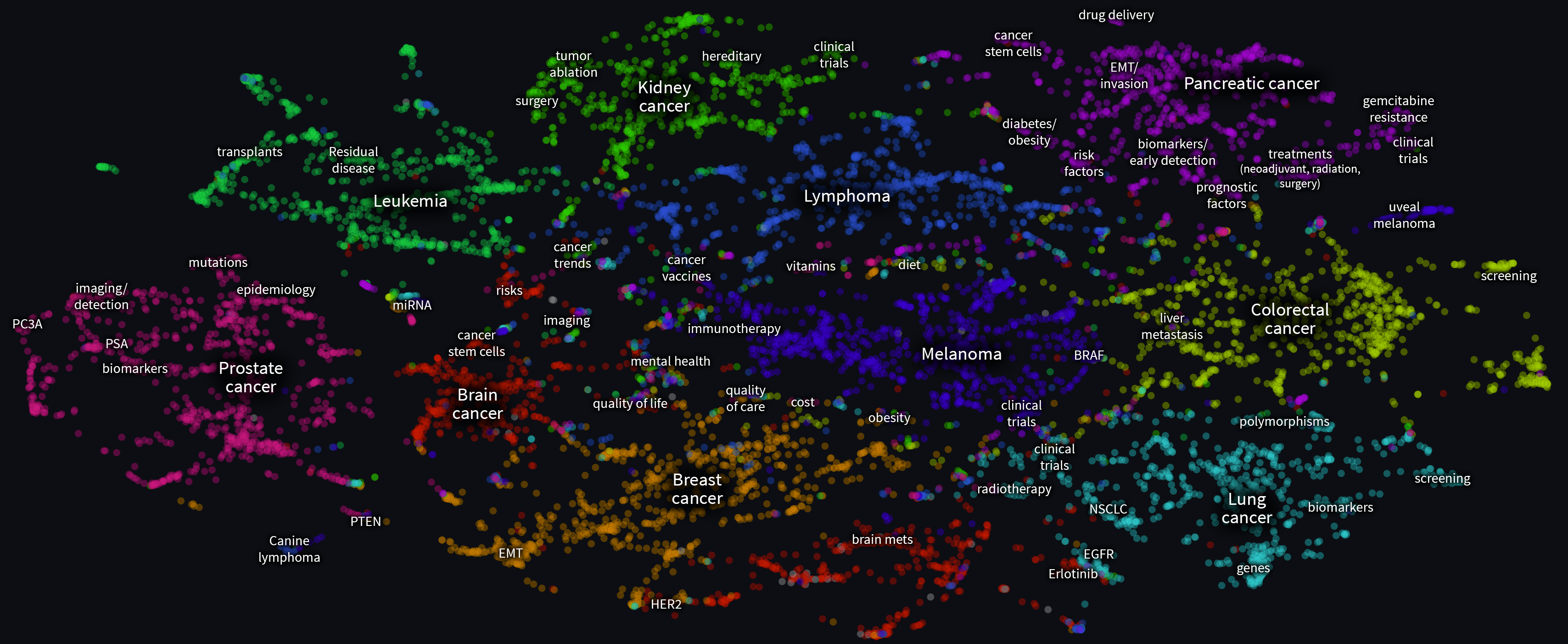}
  \caption{Topic clusters found in the 2D embedding.}
  \label{fig:vizlabels}
\end{figure*}

When computing the probability distribution for the input data, a perplexity parameter has to be set, which relates to the expected number of nearest neighbors of a data point. Additionally, the dimensionality and sparseness of the input vectors and resulting pairwise euclidean distances influence the solution. We experimented with different values and set the perplexity to 15 and the number of LSA components to 150, which provides a good trade-off between grouping together articles from the same field as well as preserving subclusters for specific topics. To verify the quality of the embedding, a postdoc in oncology examined the visualization and identified the topics of papers belonging to the same cluster (Figure~\ref{fig:vizlabels}). Besides topics related to a specific cancer type, it was also possible to identify interdisciplinary clusters, e.g.~concerning the quality of life or the influence of certain diets on the development of cancer, which supports the idea that valuable insights might be gained from broadening the literature research beyond one's usual focus.

\section{Search by similarity}\label{sec:search}
In addition to a simple keyword search, PubVis offers the possibility to match a whole abstract against the database of papers. This can be helpful, for example, when drafting a new paper, to quickly verify that no recently published papers central to your case were overlooked. For the search to work, first an index has to be created, linking every word occurring in any of the collected texts to the documents it occurs in. This can easily be accomplished by transposing the documents' tf-idf BOW vectors, which additionally ensures that frequent but meaningless words do not have a significant influence on the search results. When a new abstract is then submitted for search, the set of all words occurring in it is used to access the index and the scores associated with all matching articles are aggregated to yield the final results (Figure~\ref{fig:search} in the Appendix). The submitted abstracts are only used to search for related articles and not stored in the database.

\section{Personalized recommendations}\label{sec:rec}
\begin{figure}
  \centering
      \includegraphics[width=0.49\textwidth]{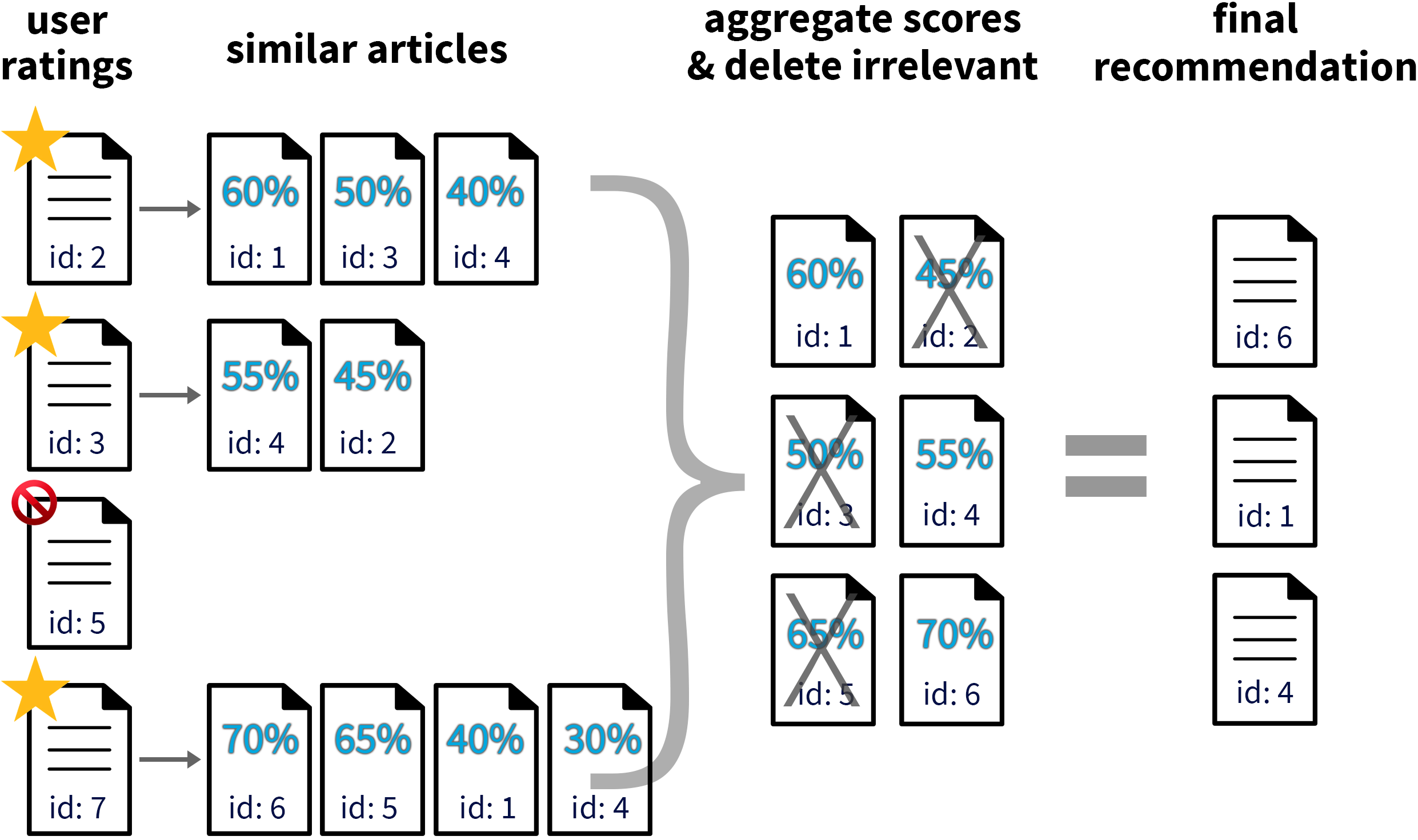}
  \caption{Process of generating personalized content based article recommendations.}
  \label{fig:rec}
\end{figure}

For every paper in the app, a user can indicate whether it is relevant or irrelevant for him or her. Based on this collected information, personalized content based article recommendations can be generated for each user (Figure~\ref{fig:rec}). 
For all articles the user has marked as relevant, similar articles are retrieved. These articles are then combined in a set of potential suggestions and labeled with the maximum similarity score each article received from one of the relevant articles.\footnote{Taking the maximum and not the average of the similarity scores for an article associated with multiple papers a user had marked as relevant accounts for the fact that users might be interested in different topics that do not necessarily produce articles related amongst each other.} From this list of potential suggestions, all articles are then removed that the user has previously marked as irrelevant. These negative ratings are not taken into account when generating the initial list of suggestions, however, as this does not improve the quality of such content based recommendations \cite{achakulvisut2016science}.

For the PubVis instance running online, the preferences of multiple users are tracked by identifying them via cookies.

\section{Discussion and Outlook}
With an elaborate set of features, the PubVis app can aid researchers in the exploration and discovery of scientific publications. With an interactive visualization of a large collection of articles, the user can quickly obtain an overview of a field. Additionally, personalized article recommendations, an easy to navigate network of similar papers, and advanced search functionalities effectively provide users with relevant content.
The open source app is easy to set up and run locally on a desktop computer and by using the built-in interface to the PubMed and arXiv APIs, the articles displayed in PubVis can be tailored to the user's interests.

While, with a reasonable amount of RAM (8-16GB) available, the app can easily cope with more than $10,000$ articles, scalability is certainly an issue. Running the app itself is less problematic, but the content updates in the current setup, especially computing the article similarities and the two dimensional embedding with t-SNE (both $\mathcal{O}(n^2)$) can quickly exceed the available resources. If more papers should be included in the app, it might be necessary to switch to iterative updates and the more efficient (but less exact) Barnes-Hut implementation of t-SNE ($\mathcal{O}(n\log{}n)$) \cite{maaten2013barnes}. However, since it can be assumed that individual users are generally only interested in well constrained research areas and it is enough to execute these content updates only a few times per month, for average use cases the current performance should be sufficient. We see this app as a big step towards making literature research simpler and more enjoyable.

Further development of PubVis will focus on making the app even easier to set up (i.e.~not requiring the usage of a terminal to run it), providing access to other journals and conference abstracts via custom web scrapers, and using the app for other types of content, e.g.~to complement movie recommendation systems.

\section*{Acknowledgments}
I would like to thank Dmitry Monin for programming the PubVis front-end, Alice Nomura for annotating the topic clusters in the PubVis visualization, and Antje Relitz for her helpful comments on the manuscript draft.

\bibliography{../../phd_collected.bib}
\bibliographystyle{plainnat}
\clearpage

\onecolumn
\section*{Appendix}

\begin{figure*}[h!]
  \centering
      \includegraphics[width=0.95\textwidth]{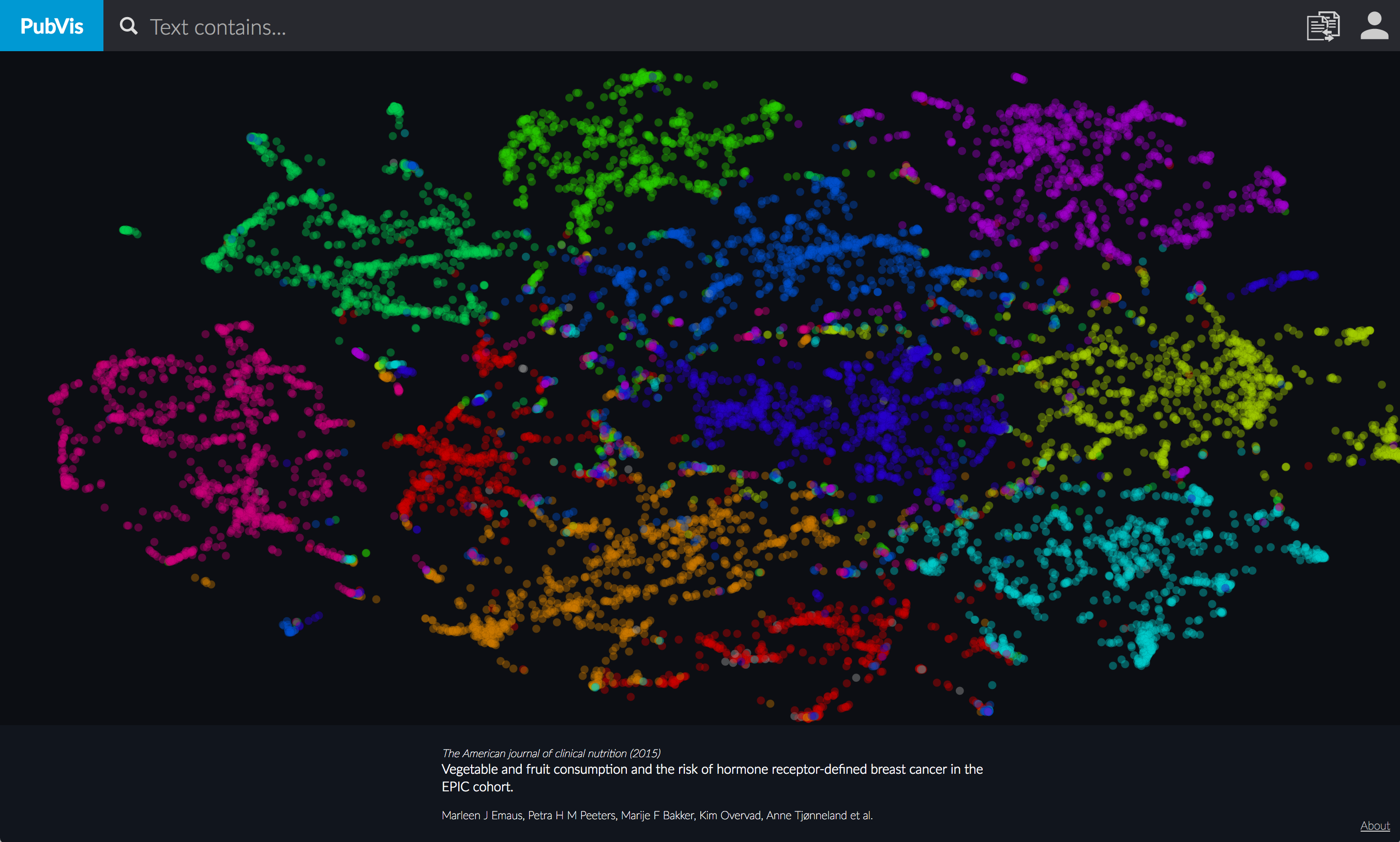}
  \caption{Interactive visualization of a collection of papers.}
  \label{fig:viz}
\end{figure*}
\begin{figure*}[h!]
  \centering
      \includegraphics[width=0.95\textwidth]{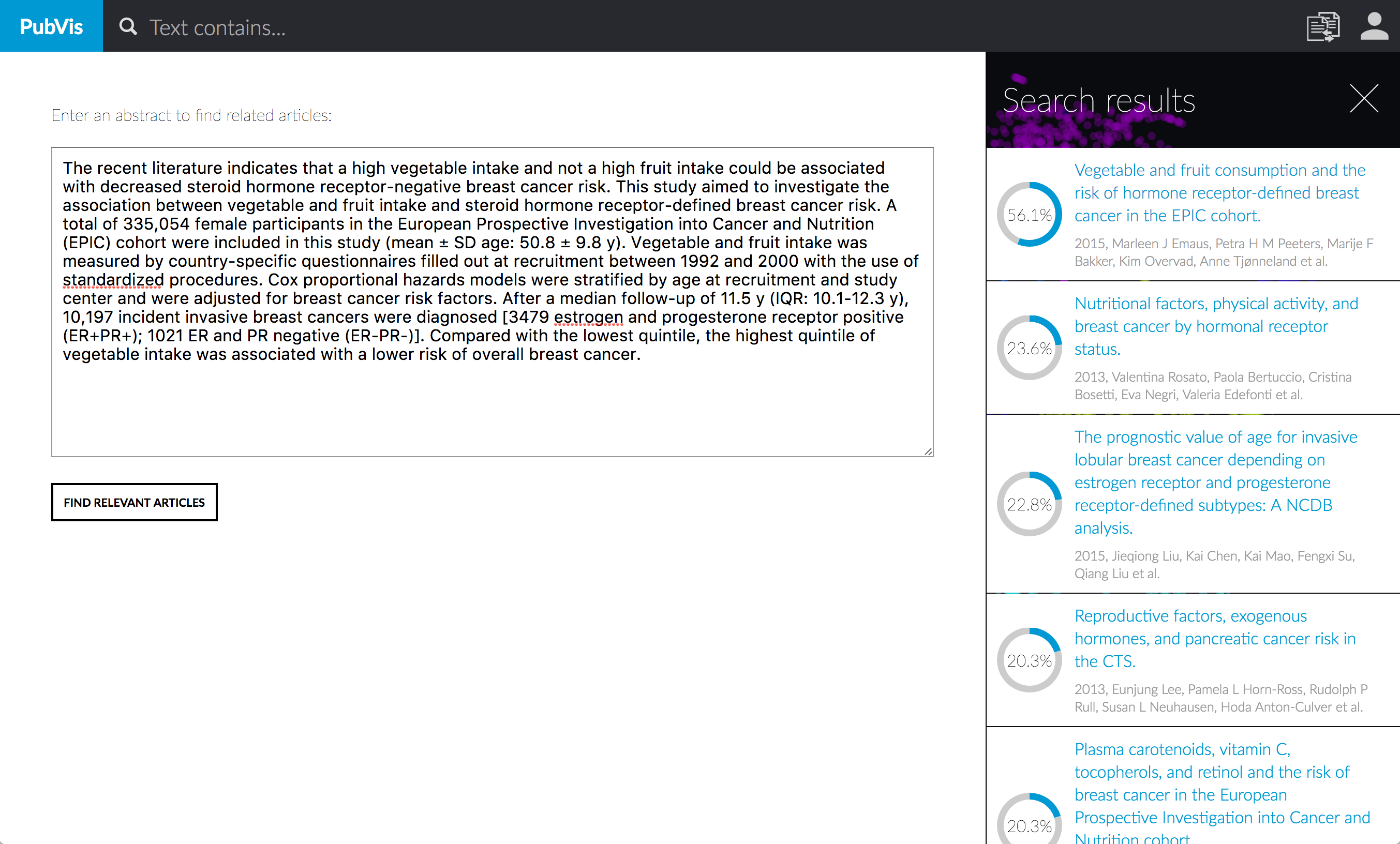}
  \caption{Search results based on a submitted abstract.}
  \label{fig:search}
\end{figure*}

\end{document}